\documentclass [aps, prx, reprint, floatfix, superscriptaddress]{revtex4-1}%
\usepackage{graphicx}
\usepackage{bm,amsmath,amssymb,mathrsfs,dcolumn}
\usepackage [colorlinks=true,linkcolor=blue,citecolor=blue,urlcolor=blue,linktoc=page,bookmarks=false,pdfstartview={FitH},pdfborder={0 0 0.0 [3 3]}]{hyperref}
\usepackage [usenames]{xcolor}
\usepackage{lipsum}
\bibpunct{[}{]}{,}{n}{}{}
\usepackage {subfigure}

\begin{document}
\title{Magnon hybridization in ferrimagnetic heterostructures}
\author {Song Li}
\affiliation {School of Science, Tianjin University, Tianjin 300072, China}
\affiliation {The Center for Advanced Quantum Studies and Department of Physics, Beijing Normal University, Beijing 100875, China}
\author {Ka Shen}
\email{kashen@bnu.edu.cn}
\affiliation {The Center for Advanced Quantum Studies and Department of Physics, Beijing Normal University, Beijing 100875, China}
\author {Ke Xia}
\affiliation {Beijing Computational Science Research Center, Beijing 100193, China}
\date{\today }
\pacs{72.25.-b, 73.50.lw, 72.10.Bg}

\begin{abstract}
  We study magnon hybridization in a ferrimagnetic heterostructure consisting of ultrathin gadolinium iron garnet and yttrium iron garnet layers and show the localized and extended spatial profiles of the magnon modes with different polarizations. These modes are expected to have distinct thermal excitation properties in the presence of a temperature gradient across the heterostructure. From a quantitative analysis of their consequences on longitudinal spin Seebeck effect, we predict an observable shift of the sign-changing temperature with respect to the one previously observed in gadolinium iron garnet. Moreover, the sign-changing point of spin Seebeck signal is found to be tunable by YIG thickness. Our results suggest the necessity of taking into account the temperature difference between the magnon modes in ferrimagnetic heterostructures.
\end{abstract}
\maketitle
\section{Introduction}
Magnons~\cite{hp:pr1940,blochmganon:1930}, collective excitations in magnetic ordering systems, have been considered as potential information carriers in low-power devices. They can be activated by microwave~\cite{Yuspinwaveprl:2019,liu2018long}, laser~\cite{kirilyuklaserrmp:2010,satoh2012directional,shenlaserprl:2015} and thermal fluctuation~\cite{uchida2008observation,uchida2010spin,uchidalongiapl"2010} and interact with each other through, e.g., exchange coupling~\cite{akhiezer1968spin,kashenpurespin2019:prb} and magnetic dipolar interaction~\cite{akhiezer1968spin,Serga_2010,shendipolarprl:2020}.
In the past decade, many interesting magnon-related phenomena, such as spin Seebeck effect (SSE)~\cite{uchida2008observation,Uchida:naturematerials2010}, orbital Nernst effect of magnon~\cite{zhang2019orbital} and corner states in ferromagnetic breathing Kagome lattice~\cite{Sil_2020},
 have been reported. To understand the underlying physics of these phenomena, we need to explore the properties of magnons.

While earlier studies mainly focused on the magnons in a single magnetic layer, recent experiments revealed attractive properties in hybrid magnetic structure. For instance, unexpected enhancements were observed in spin pumping (SP)~\cite{wangnioprl:2014} and SSE signals~\cite{Linnioprl:2016} when an ultrathin antiferromagnetic NiO was inserted between yttrium iron garnet (YIG) and Pt layers ~\cite{chentemprb:2016,Rezendenioprb:2016,Khymynprb:2016,tataraafprb:2019}. Such enhancements were attributed to either the increased interfacial spin mixing conductance~\cite{chentemprb:2016,Rezendenioprb:2016} or the interference of evanescent waves~\cite{Khymynprb:2016}, one type of hybrid spin waves in the magnetic bilayer structure. Recent phase-resolved x-ray pump-probe measurements showed the evidence of magnon transmission via the evanescent waves~\cite{Dabrowskicoprl:2020}. Another important feature in hybrid magnetic structure is the anticrossing between different ferromagnetic resonances observed in, e.g., YIG-Ni~\cite{chenyigniprl:2018}, YIG-Co~\cite{klingleryigcoprl:2018} and YIG-CoFeB~\cite{Qinygcfosr:2018} bilayer structures, which reveals the formation of hybrid spin wave modes around the anticrossing. These mode hybridizations can affect the measurable quantities in practical experiments. For example, the suppression and enhancement of ferromagnetic resonance linewidth were observed for the in-phase and out-of-phase coupled modes, respectively, in YIG-permalloy (YIG-Py) system~\cite{LiYIGPyprl:2020}. A precise description of these observations requires a detailed calculation of the hybrid magnon modes which could play an essential role, especially when part of the system is of only a few nanometers thick.

YIG as one of the most important magnetic materials due to its low-damping coefficient is usually grown on gadolinium gallium garnet (GGG) substrate~\cite{Dubs_2017}. Recently, Gomez-Perez et al. showed that near the interface between YIG and GGG, the Gd atoms from GGG can diffuse into the YIG layer and substitute Y atoms in YIG, forming a natural YIG-GdIG magnetic bilayer~\cite{felixyiggdigprap:2018}, where the thickness of the GdIG layer is around 3 nanometers. Theoretically, whereas the magnon spectra in both YIG and GdIG have been studied in literatures~\cite{harris:pr1963,shen:jnp2018,shengdigprb:2019,xieyigprb:2017,wanglianweiprb:2020}, the hybrid magnon spectrum in their hybrid system is still missing. On the other hand, although GdIG shares the same structure with YIG, its SSE~\cite{geprags:nncomm2016} is found to be quite different from that in YIG~\cite{Rezendeprb:2014,uchida2010spin}. An interesting question one may ask is: What are the consequences of hybrid magnon modes in YIG-GdIG bilayer in the spin Seebeck measurement. Therefore in this work, we calculate the hybrid magnon spectrum in YIG-GdIG bilayer system and analyze its consequences in the longitudinal spin Seebeck effect (LSSE).

\section{Hybrid Spectrum and Mode Hybridizations in the YIG-GdIG bilayer system}\label{sec2}
\subsection{Qualitative analysis}
 \begin{figure} [t]
\includegraphics [width=8.5cm]{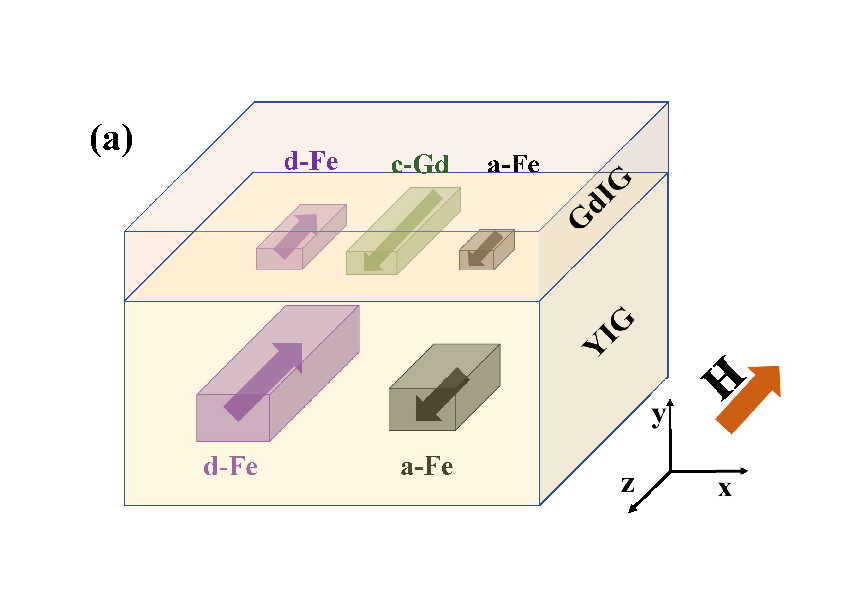}
\includegraphics [width=8.5cm]{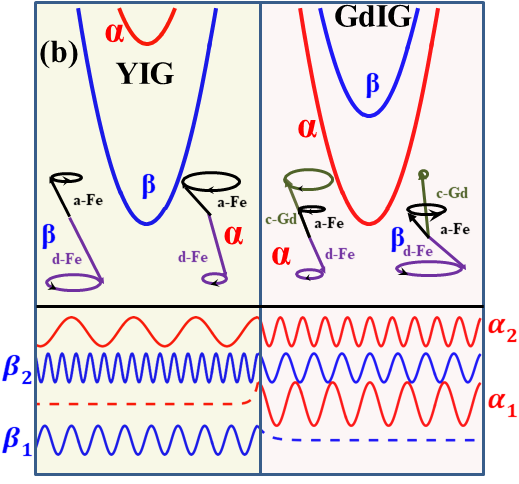}
\caption{
Schematic of the structure (a), the magnon spectrum and hybrid modes (b) in the YIG-GdIG hybrid system, where the red and blue lines stand for the left-handed ($\alpha$) and right-handed ($\beta$) modes, respectively.}
\label{fig1}%
\end{figure}
We consider the situation with the thickness of YIG in the YIG-GdIG bilayer much larger than that of GdIG. As experimentally demonstrated in Ref.\cite{felixyiggdigprap:2018}, the net magnetic moments of the two parts in such structure align antiparallelly under a weak in-plane field. Therefore, as sketched in Fig.\ref{fig1}(a), when a small magnetic field is applied along $-\hat{\bf z}$, the magnetic moment of the YIG layer dominated by the d-Fe sublattice~\cite{harris:pr1963} points to $-\hat{\bf z}$ and that of the GdIG layer determined by c-Gd sublattice~\cite{harris:pr1963} points to $\hat{\bf z}$. Notice also that both orientations of a-Fe and d-Fe sublattice in the GdIG layer are the same as those in the YIG layer.

In such a system, the YIG layer contains two types of modes, where the one with lower frequency (higher frequency) is dominated by the precession of d-Fe (a-Fe) sublattice~\cite{shen:jnp2018} and the GdIG layer contains three types of magnons, dominated by the precessions of c-Gd sublattice, d-Fe sublattice and a-Fe sublattice, respectively~\cite{shengdigprb:2019}. The magnon dispersions in the two parts of this bilayer system are sketched in the upper panel of Fig.\ref{fig1}(b), where the two bands in the two layers are of opposite chiralities with the gap in YIG larger than that in GdIG~\cite{harris:pr1963,shen:jnp2018,wanglianweiprb:2020}. Notice that the high energy mode dominated by the precession of a-Fe sublattice in GdIG will not affect our main results, we therefore discard it in the figure. Due to the antiferromagnetically aligned magnetic moments of Gd atoms and d-Fe atoms (as seen in Fig.\ref{fig1}(a)), the lower branches in the YIG layer and the GdIG layer carry opposite spin angular momentums. By taking into account the hybridization between the two layers, one expects four types of hybrid modes (as plotted in the lower panel of Fig.\ref{fig1}(b)): left-handed ($\alpha_1$) modes propagating in the GdIG layer but evanescent in the YIG layer and right-handed ($\beta_1$) modes propagating in the YIG layer but evanescent in the GdIG layer, right-handed ($\beta_2$) and left-handed ($\alpha_2$) modes propagating in both layers.
\begin{figure*}[htpb] 
\includegraphics[width=2.3in]{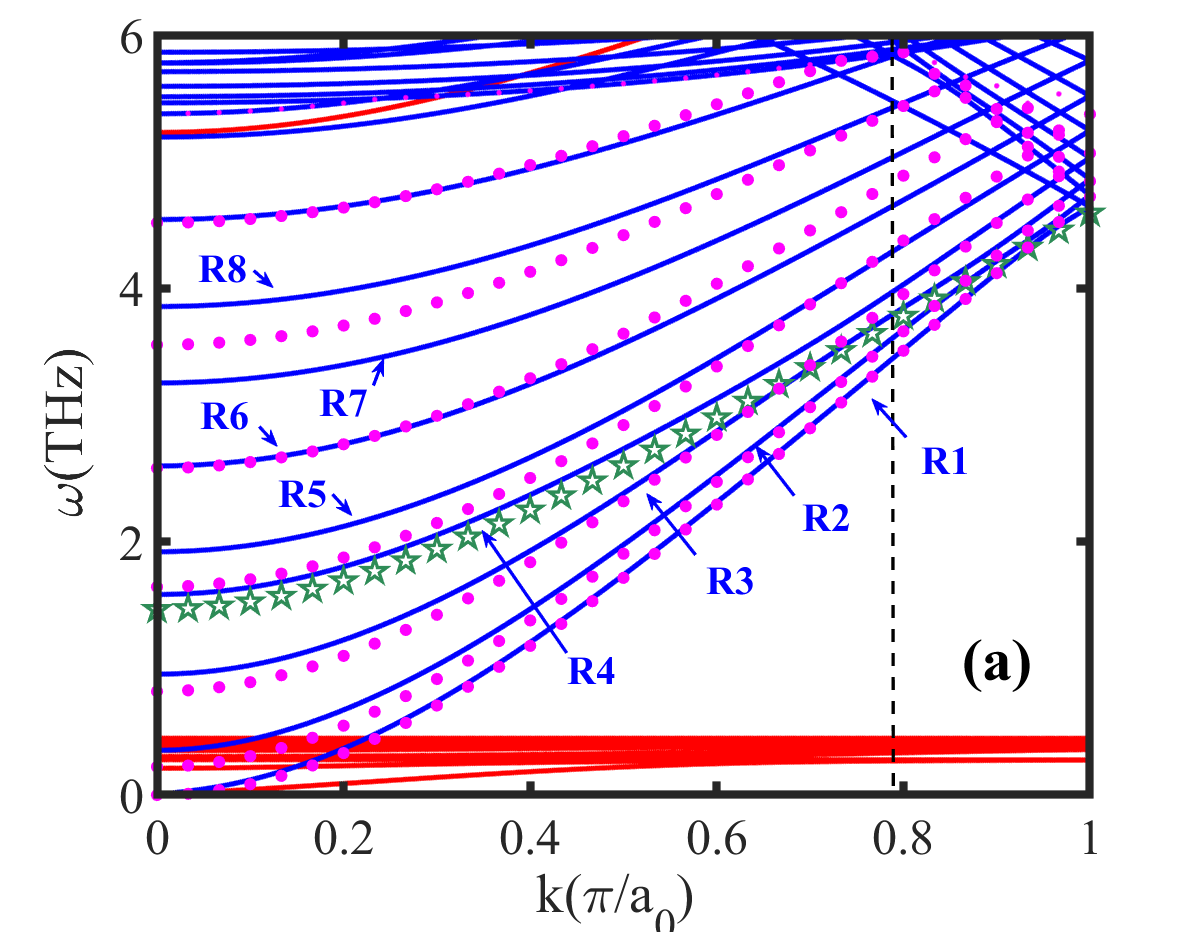}
\includegraphics[width=2.3in]{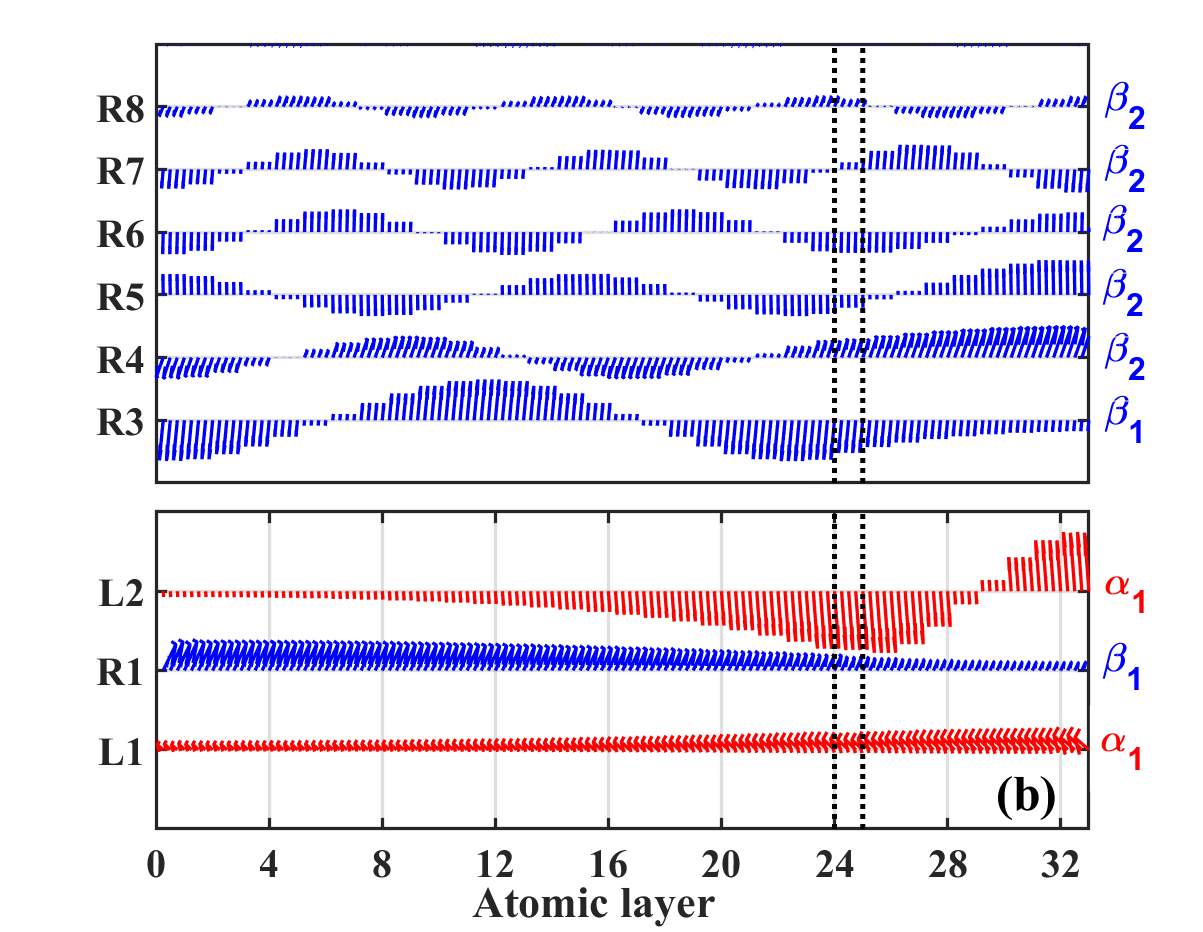}
\includegraphics[width=2.3in]{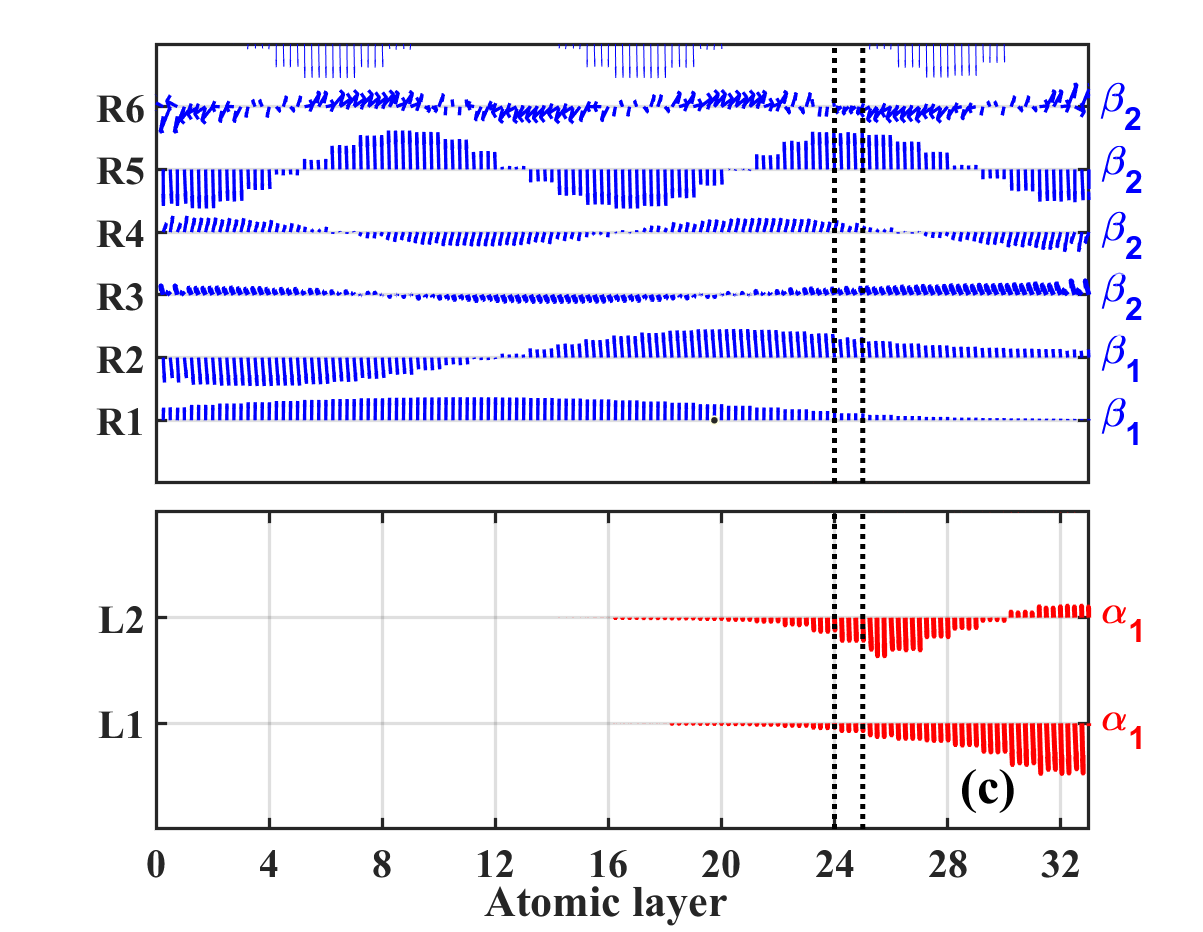}
\caption{
(a) Spin wave dispersion in YIG(7.4nm)-GdIG(2.5nm) [001] bilayer system, where red and blue lines are specified for $\alpha$ and $\beta$ modes, respectively. Pink dots and green pentagrams are $\beta$ branches in 7.4-nm-thick YIG and 2.5-nm-thick GdIG. (b) and (c) Transverse spin orientations of a-Fe atoms along the bilayer at $k=0.05\pi/a_0$ and $k=0.78\pi/a_0$. Black dotted lines enclose the interfacial regions between YIG and GdIG. The capital letters, L and R, are the abbreviations for left-handed mode and right-handed mode, respectively.}
\label{fig2}
\end{figure*}
\subsection{Heisenberg model}
To calculate the concrete spectrum, we apply the atomic spin exchange model to our bilayer structure
\begin{align}
 H	&=	-\text{\ensuremath{\sum_{n=1}^{N}[\sum_{i=1}^{n_{a}}\sum_{\left|{\bf r}_{ij}\right|=r_{aa}}J_{ij}^{aa}S_{a}({\bf R}_{in})\cdot S_{a}({\bf R}_{in}+{\bf r}_{ij})}} \notag \\
 		&+\sum_{i=1}^{n_{d}}\sum_{\left|{\bf r}_{ij}\right|=r_{dd}}J_{ij}^{dd}S_{d}({\bf R}_{in})\cdot S_{d}({\bf R}_{in}+{\bf r}_{ij}) \notag \\
 		&+\sum_{i=1}^{n_{c}}\sum_{\left|{\bf r}_{ij}\right|=r_{cc}}J_{ij}^{cc}S_{c}({\bf R}_{in})\cdot S_{c}({\bf R}_{in}+{\bf r}_{ij}) \notag \\
		&+2\sum_{i=1}^{n_{a}}\sum_{\left|{\bf r}_{ij}\right|=r_{ad}}J_{ij}^{ad}S_{a}({\bf R}_{in})\cdot S_{d}({\bf R}_{in}+{\bf r}_{ij}) \notag \\
		&+2\sum_{i=1}^{n_{a}}\sum_{\left|{\bf r}_{ij}\right|=r_{ac}}J_{ij}^{ac}S_{a}({\bf R}_{in})\cdot S_{c}({\bf R}_{in}+{\bf r}_{ij}) \notag \\
		&+2\sum_{i=1}^{n_{d}}\sum_{\left|{\bf r}_{ij}\right|=r_{dc}}J_{ij}^{dc}S_{d}({\bf R}_{in})\cdot S_{c}({\bf R}_{in}+{\bf r}_{ij})],
 \end{align}where $n_a$, $n_c$ and $n_d$ are the total numbers of local spins at a-Fe, c-Gd and d-Fe sites in one unit cell. $r_{ss'}$ and $J^{ss'}_{ij}$ are the nearest neighbor distance and the position dependent exchange coupling between magnetic atoms $s$ and $s'$. From crystal structure of garnet, we extract the set of nearest neighbor distances as $r_{aa}=(\sqrt{3}/4)a_0$, $r_{dd}=(\sqrt{6}/8)a_0$, $r_{dc}=(1/4)a_0$ and $r_{ad}=r_{ac}=(\sqrt{5}/8)a_0$, with $a_0=1.24$ nm. We then derive the bosonic Bogoliubov-de Gennes (BdG) Hamiltonian as~\cite{shen:jnp2018,shengdigprb:2019}
\begin{align}\label{bosonic}
  H_{\bf k}&=\sum_{i,j=1}^{n_a}a_{i}^{\dagger}({\bf k})A_{ij}({\bf k})a_{j}({\bf k})+\sum_{i,j=1}^{n_c}c_{i}^{\dagger}({\bf k})C_{ij}({\bf k})c_{j}({\bf k})\notag\\
  &+\sum_{i,j=1}^{n_d}d_{i}^{\dagger}(-{\bf k})D_{ij}(-{\bf k})d_{j}(-{\bf k})\notag\\
  &+\sum_{i=1}^{n_a}\sum_{j=1}^{n_c}\left[a_{i}^{\dagger}({\bf k})B^{ac}_{ij}({\bf k})c_{j}({\bf k})+h.c.\right]\notag\\
  &+\sum_{i=1}^{n_a}\sum_{j=1}^{n_d}\left[a_{i}^{\dagger}({\bf k})B^{ad}_{ij}({\bf k})d_{j}^{\dagger}(-{\bf k})+h.c.\right]\notag\\
  &+\sum_{i=1}^{n_c}\sum_{j=1}^{n_d}\left[c_{i}^{\dagger}({\bf k})B^{cd}_{ij}({\bf k})d_{j}^{\dagger}(-{\bf k})+h.c.\right].
\end{align}The matrix elements $A_{ij}$, $C_{ij}$, $D_{ij}$ and $B_{ij}^{ss'}$ are given in Appendix \ref{parameter}. Operators $a_i$, $d_i$ and $c_i$ are defined by Holstein-Primakoff (H-P) transformation~\cite{hp:pr1940} of their atomic spins as
  \begin{align}
  S_{a,i}^z&=S_{a,i}-a^\dagger_i a_i, S_{a,i}^{+}=\sqrt{2S_{a,i}-a^\dagger_i a_i} a_i, \notag\\
  S_{c,i}^z&=S_{c,i}-c^\dagger_i c_i, S_{c,i}^{+}=\sqrt{2S_{c,i}-c^\dagger_i c_i} c_i, \notag\\
  S_{d,i}^z&=-S_{d,i}+d^\dagger_i d_i, S_{d,i}^{-}=d_i \sqrt{2S_{d,i}-d^\dagger_i d_i}.
  \end{align}
By diagonalizing the Hamiltonian through paraunitary transformation~\cite{clopa:pa1978,flebus:prb2017}, we can obtain the magnon spectrum of this hybrid bilayer system. The resultant eigenstates are linear combination of the local spin operators at a, c and d sites~\cite{shen:jnp2018}
  \begin{align}\label{bo}
  \alpha_{\bf k}^{i'}&=p_{a,{\bf k}}^{i'i}a_{i,{\bf k}}+p_{c,{\bf k}}^{i'j}c_{j,{\bf k}}+p_{d,-{\bf k}}^{i'l}d_{l,{-{\bf k}}}^{\dagger}, \notag\\
  \beta_{\bf k}^{j'}&=p_{\overline{a},-{\bf k}}^{j'i}a_{i,{-{\bf k}}}^{\dagger}+p_{\overline{c},-{\bf k}}^{j'j}c_{j,{-{\bf k}}}^{\dagger}+p_{\overline{d},{\bf k}}^{j'l}d_{l,{\bf k}},
  \end{align}
where superscripts $i'=1, \cdots, n_a+n_c$ and $j'=1, \cdots, n_d $ are the indexes of modes $\alpha$ and $\beta$. $i=1, \cdots, n_a$, $j=1, \cdots, n_c $ and $l=1, \cdots, n_d$ here are the indexes of local spin operators at a, c and d sites. Einstein summation convention is applied for $i$, $j$, and $l$.

\subsection{Numerical results}\label{sec3}
In this subsection, we present the hybrid magnon spectrum and wave functions in a YIG-GdIG [001] bilayer system with YIG and GdIG layers 6-unit-cell-thick (7.4 nm) and 2-unit-cell-thick (2.5 nm), respectively. We adopt $J_{ij}^{aa}=-0.329 {\rm meV}$, $J_{ij}^{dd}=-1.161 {\rm meV}$, $J_{ij}^{ad}=-3.449 {\rm meV}$  and $S_{a,i}=S_{d,i}=2.5$ for the YIG layer~\cite{harris:pr1963} and $J_{ij}^{aa}=-0.081 {\rm meV}$, $J_{ij}^{dd}=-0.137 {\rm meV}$, $J_{ij}^{ad}=-2.487 {\rm meV}$, $J_{ij}^{ac}=0.032 {\rm meV}$, $J_{ij}^{cd}=-0.157 {\rm meV}$, $S_{a,i}=2.1$, $S_{d,i}=2.05$ and $S_{c,i}=3.5$ for the GdIG and interfacial regions~\cite{wanglianweiprb:2020}. Fig.\ref{fig2}(a) shows the magnon spectra of the two separated layers and their hybrid bilayer system. For a 7.4-nm-thick YIG film, the lowest seven magnon branches, corresponding to the right-handed ferromagnetic resonance mode and its subbands with increasing nodes are shown in pink dots and $\alpha$ modes are also found in high-frequency range (not shown). For the magnons in a 2.5-nm-thick GdIG layer, a $\beta$ mode (shown in green pentagrams) and many $\alpha$ modes lying within 0-0.5 THz are plotted. As we can see in this figure, the $\beta$ mode in the GdIG layer crosses with other three $\beta$ modes in the YIG layer. As a result of interlayer coupling in the YIG-GdIG structure, gaps are opened at these crossing points in the hybrid magnon spectrum (shown in the red and blue lines).
The slight deviation between the bands of the bare YIG film and the hybrid system is due to the missing atomic layer at the interface.

To give more details of the hybrid modes in this bilayer system, we plot the instantaneous orientations of magnetic moments for a-Fe atoms in x-y plane along the whole bilayer system. Fig.\ref{fig2}(b) shows the low-frequency hybrid wave functions near the center of Brillouin zone ($k=0.05\pi/a_0$), including three types of hybrid modes, i.e., $\alpha_1$ modes for L1 and L2, $\beta_1$ modes for R1, R2 (not shown), R3 and $\beta_2$ modes for R4, R5, $\cdots$, R8. As the anticrossings in the hybrid spectrum reveal the formation of new hybrid modes, we plot the wave functions at $k=0.78\pi/a_0$ (marked by the dashed line in Fig.\ref{fig2}(a)) in Fig.\ref{fig2}(c) and find that R3 is transformed from $\beta_1$ mode to $\beta_2$ mode while the types of the other modes remain unchanged. Note that $\alpha_2$-type magnons are also found at these two wavevectors in the high-frequency regime (not shown here).

\section{LSSE in YIG-GdIG-NM trilayer system}\label{sec4}
\begin{figure} [hbt]
\includegraphics [width=8.6cm]{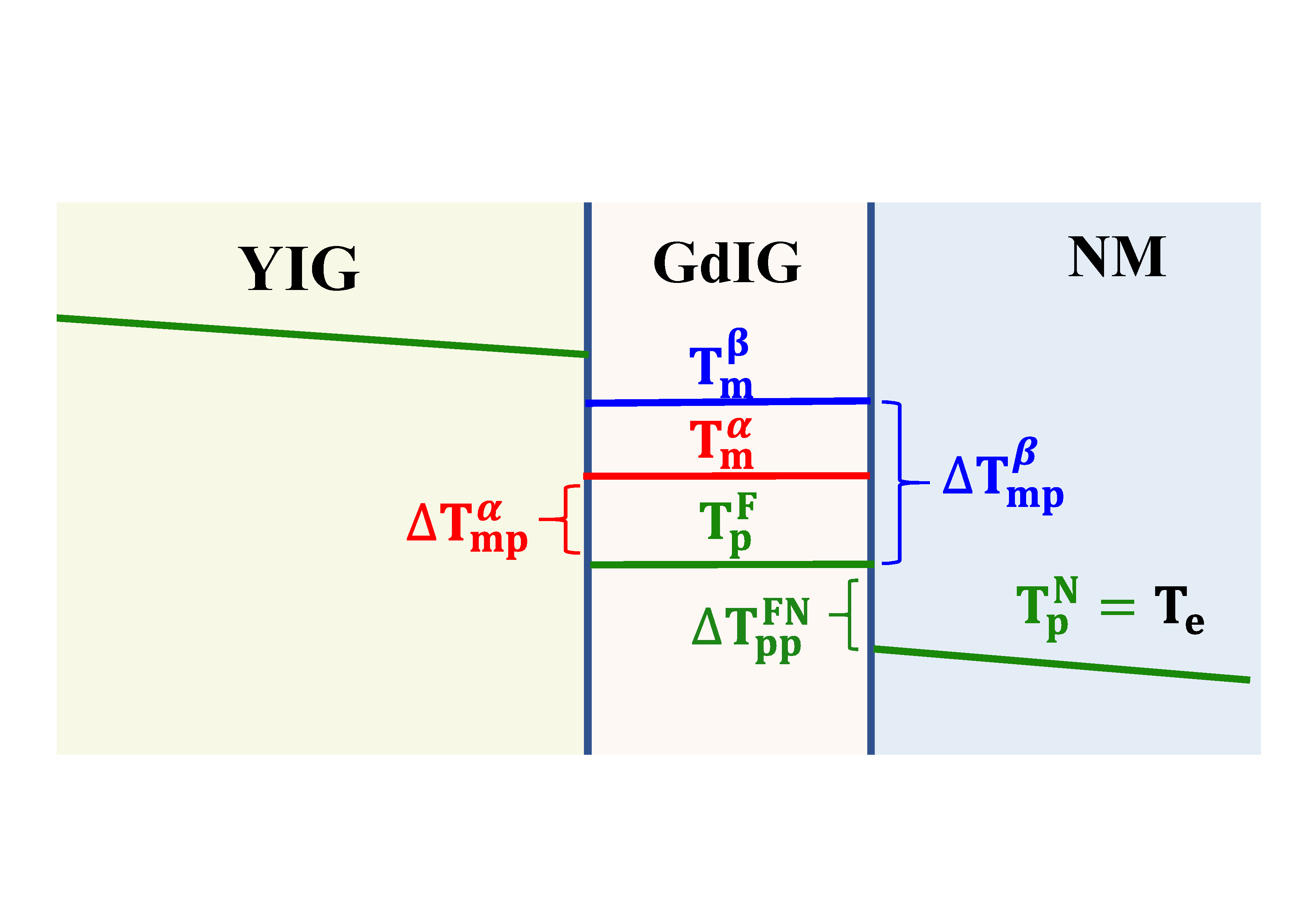}\\
\caption{Relation between the local temperature of phonons (green curve) in YIG-GdIG-NM trilayer in LSSE configuration and that of magnons for $\alpha$ (red curve) and $\beta$ (blue curve) modes in GdIG layer.
}
\label{fig3}%
\end{figure}
With the hybrid magnon spectra and wave functions, one can analyze the consequences of hybrid modes in the transport properties, e.g., LSSE, in which the nonequilibrium between the phonons and magnons near the magnetic insulator-normal metal (NM) interface is considered as the driving force~\cite{olssondtrew2018,shen:prb2016,xiao:prb2010,agrawalblsprl:2013,schreier:prb2013}. During the measurement of LSSE, the magnons accumulated at the GdIG-NM interface are mainly $\alpha_1$-type and $\beta_2$-type while the contributions from $\beta_1$-type and $\alpha_2$-type modes are far lesser due to the blockage by GdIG and the high excitation frequency, respectively. Considering the extended and localized features of the $\beta_2$-type and $\alpha_1$-type magnons, the temperatures of the $\alpha$ and $\beta$ magnons near the GdIG-NM interface in the LSSE should be different. Specifically, the differences between the temperature of $\beta$ magnons, $\rm{T}^{\beta}_m$, and the temperature of phonons, ${\rm T^F_p}$, should be larger than that between the temperature of $\alpha$ magnons, $\rm{T}^{\alpha}_m$, and ${\rm T^F_p}$. Therefore, we introduce a two-temperature model to describe the LSSE in the YIG-GdIG-NM trilayer system as sketched in Fig.\ref{fig3}, where an interfacial temperature discontinuity between phonons in the GdIG layer and the NM layer, equal to the temperature of electrons, ${\rm T_e}$, is also introduced. Note that since the thickness of each magnetic layer is smaller than the phonon mean free path, we here assume the local temperature inside the YIG and GdIG layers are both uniform and focus on the temperature difference across each interface due to Kapitza heat resistance.

Within the linear-response regime, the spin currents generated in YIG-GdIG-NM trilayer are proportional to the temperature differences between magnons and electrons~\cite{shen:prb2016,Ohnumaferriprb:2013,geprags:nncomm2016}, i.e.,
\begin{align}\label{tmod}
I_s=A_{\rm T}{\rm \Delta T^{\alpha}_{me}}-B_{\rm T}{\rm \Delta T^{\beta}_{me}},
\end{align} where $A_{\rm T}$ and $B_{\rm T}$ are the spin Seebeck conductances (SSC) for $\alpha$ magnons and $\beta$ magnons. The temperature differences in Eq.(\ref{tmod}) can be read from Fig.\ref{fig3}
\begin{align}
{\rm \Delta T^{\alpha}_{me}} &={\rm \Delta T^{\alpha}_{mp}}+{\rm \Delta T^{FN}_{pp}},\notag \\
{\rm \Delta T^{\beta}_{me}} &={\rm \Delta T^{\beta}_{mp}}+{\rm \Delta T^{FN}_{pp}},
\end{align} where ${\rm \Delta T^{\alpha}_{mp}}$, ${\rm \Delta T^{\beta}_{mp}}$ are the temperature differences between $\alpha$ modes and local phonons, $\beta$ modes and local phonons near the GdIG-NM interface. Based on Eq.(\ref{tmod}), we can compare the situations in the conventional GdIG-NM bilayer and YIG-GdIG-NM trilayer: In the GdIG-NM bilayer, where ${\rm \Delta T^{\alpha}_{me}}$ and ${\rm \Delta T^{\beta}_{me}}$ are equal and fixed by the boundary, changing SSC is the only approach to tune the spin Seebeck signals; In contrast, the hybrid modes in the YIG-GdIG-NM trilayer system would cause a difference between ${\rm \Delta T^{\alpha}_{mp}}$ and ${\rm \Delta T^{\beta}_{mp}}$ and thus provide an additional possibility to manipulate LSSE.
\subsection{SSC in the LSSE}
To calculate the SSC in Eq.(\ref{tmod}), we use the s-d exchange model at the GdIG-NM interface~\cite{shen:prb2016}
   \begin{equation}\label{Hint}
   \mathcal{H'}=l^2_0\sum_{n=1}^{\mathcal{N}}\sum_{i,j\in {\rm int}}\mathcal{J}_s{\boldsymbol{S}}_{in}\cdot{\boldsymbol{\sigma}_j}\delta({\bf r}_{j}-{\bf R}_{in}),
   \end{equation}
where ${\boldsymbol{\sigma}_j}$ is the electron spin at position ${\bf r}_j$, $\mathcal{N}$ is the total number of unit cells in the 2-dimensional plane, $l_0^2$ is the area of cross section, int is the abbreviation of interface and $\mathcal{J}_s$ is the coupling strength between magnetic atoms and s electrons. One has the second-quantized Hamiltonian after performing H-P transformation~\cite{hp:pr1940} as
  \begin{align}\label{ham}
  \mathcal{H'}&=\tilde{\mathcal{J}}_a\sum_{\bf q, k}{\Big[}
  (\sum_{i\in {\rm int}}a_{i,{\bf k}}g_{\downarrow,\bf q}^{e,\dagger}g_{\uparrow,\bf q-k}^{e}+h.c.) \notag\\
  &+(\sum_{j\in {\rm int}}\eta c_{j,{\bf k}}g_{\downarrow,\bf q}^{e,\dagger}g_{\uparrow,\bf q-k}^{e}+h.c.)\notag\\
  &+(\sum_{l\in {\rm int}}d_{l,{-\bf k}}^{\dagger}g_{\downarrow,\bf q}^{e,\dagger}g_{\uparrow,\bf q-k}^{e}+h.c.){\Big]},
  \end{align}
where $g^{e}_{\uparrow(\downarrow)}$ is the annihilation operator of spin-up (down) electrons and $\tilde{\mathcal{J}}_s=\frac{1}{2}\mathcal{J}_{s}\sqrt{2\rm{S}_{s}\mathcal{N}}$. We also define $\eta=\tilde{\mathcal{J}}_c/\tilde{\mathcal{J}}_a$. After substituting the inverse transformation of Eq.(\ref{bo}),
  \begin{align}
  a_{i,{\bf k}}	&=	\sum_{i'}T_{\alpha,\bf k}^{ii'}\alpha_{\bf k}^{i'}+\sum_{j'}T_{\beta,\bf k}^{ij'}(\beta_{-\bf k}^{j'})^{\dagger},\notag \\
  c_{j,{\bf k}}	&=	\sum_{i'}T_{\alpha,\bf k}^{ji'}\alpha_{\bf k}^{i'}+\sum_{j'}T_{\beta,\bf k}^{jj'}(\beta_{-{\bf k}}^{j'})^{\dagger},\notag \\
  d_{l,{\bf k}}	&=	\sum_{i'}T_{\overline{\alpha},\bf k}^{li'}(\alpha_{-\bf k}^{i'})^{\dagger}+\sum_{j'}T_{\overline{\beta},\bf k}^{lj'}\beta_{\bf k}^{j'},
  \end{align}
into Hamiltonian (\ref{ham}), we obtain a perturbation Hamiltonian\cite{kashenpurespin2019:prb}
  \begin{align}\label{inth}
  \mathcal{H'}&=\tilde{\mathcal{J}}_a\sum_{\bf q, k}{\Big\{}
  (\sum_{i'}\mathcal{T}^{i'}_{\alpha,\bf k}\alpha_{\bf k}^{i'}g_{\downarrow,\bf q}^{e,\dagger}g_{\uparrow,\bf q-k}^{e}+h.c.)\notag \\
  &+[\sum_{j'}\mathcal{T}^{j'}_{\beta,\bf k}(\beta_{-\bf k}^{j'})^{\dagger}g_{\downarrow,\bf q}^{e,\dagger}g_{\uparrow,\bf q-k}^{e}+h.c.]
  {\Big\}}.
  \end{align}
 The coefficients are defined as
  \begin{align}\label{Tint}
  \mathcal{T}^{i'}_{\alpha,\bf k}=\sum_{i\in {\rm int}}T_{\alpha,\bf k}^{ii'}+\sum_{j\in {\rm int}}\eta T_{\alpha,\bf k}^{ji'}+\sum_{l\in {\rm int}}(T_{\overline{\alpha},-\bf k}^{li'})^{*},\notag \\
  \mathcal{T}^{j'}_{\beta,\bf k}=\sum_{i\in {\rm int}}T_{\beta,\bf k}^{ij'}+\sum_{j\in {\rm int}}\eta T_{\beta,\bf k}^{jj'}+\sum_{l \in {\rm int}}(T_{\overline{\beta},-\bf k}^{lj'})^{*}.
  \end{align}

We then follow the procedures presented in the Appendix \ref{appx} and obtain the expressions for SSCs in Eq.(\ref{tmod})
\begin{align}\label{ssc1}
A_{\rm T}&=\mathcal{D}\hbar\sum_{{\bf k}}\sum_{i'}(\partial_{{T}}n|_{T={\rm T_{eq}}})|\mathcal{T}_{\alpha,{\bf k}}^{i'}|^{2}\omega_{i'}^{{\bf k}},  \notag \\
B_{\rm T}&=\mathcal{D}\hbar\sum_{{\bf k}}\sum_{j'}(\partial_{{T}}n|_{T={\rm T_{eq}}})|\mathcal{T}_{\beta,{\bf k}}^{j',\dagger}|^{2}\omega_{j'}^{{\bf -k}},
\end{align}where $\mathcal{D}$, $n$ and ${\rm T_{eq}}$ are dimensionless coefficient defined in Appendix \ref{appx}, magnon distribution function and equilibrium temperature, respectively. From Eq.(\ref{ssc1}), we find that the magnon occupation, the dispersion of hybrid modes and the rescaled numbers of magnons accumulated at the magnetic insultor-NM interface (the square of the coefficients in Eq.(\ref{Tint})) together determine these SSCs.

\subsection{Numerical results in the YIG-GdIG-NM system}
\begin{figure} [hbt]
\includegraphics [width=8.6cm]{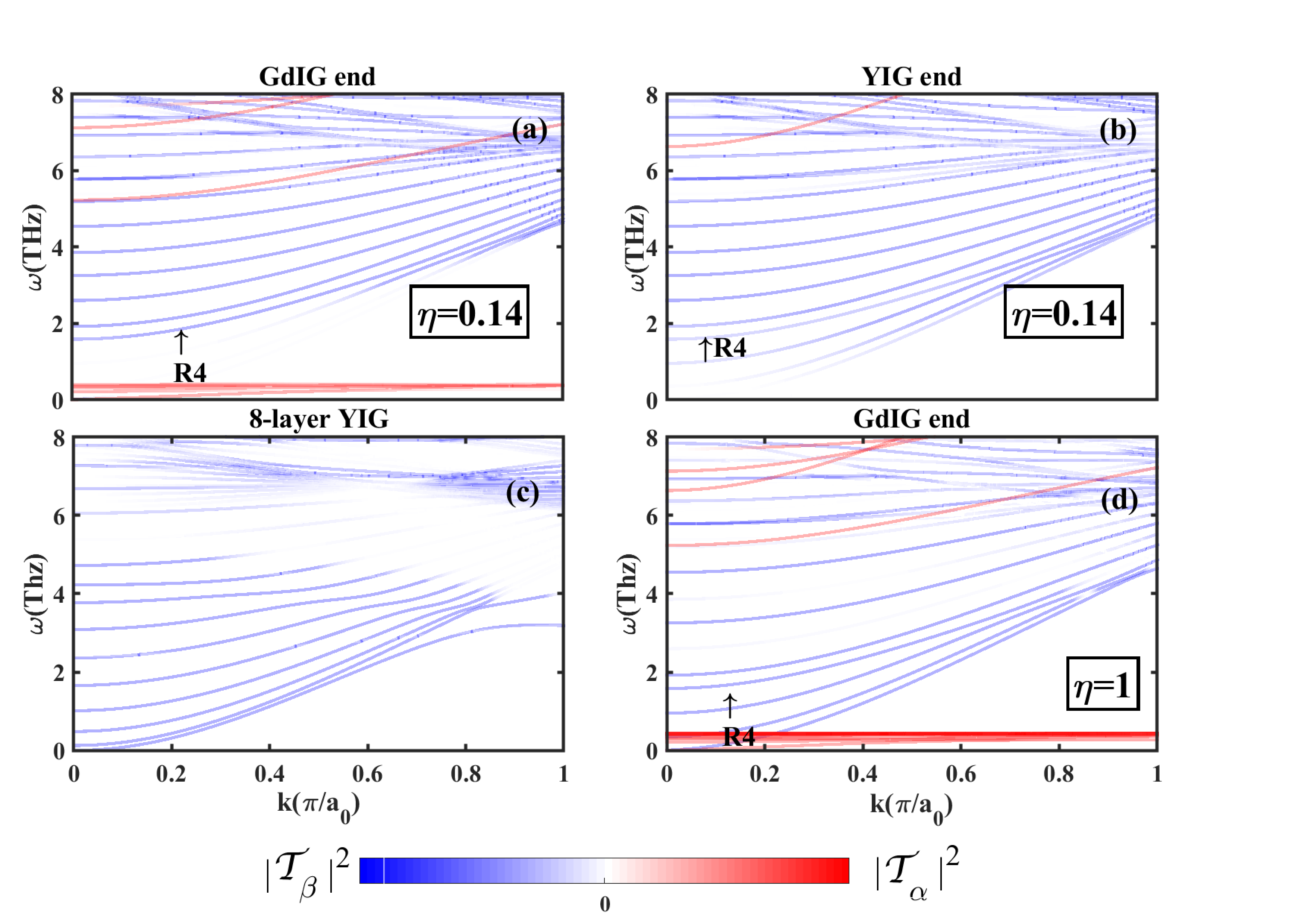}\\
\caption{Magnon spectrum weighted by the projection of wave functions at (a) the GdIG end and (b) the YIG end in YIG (7.4 nm)-GdIG (2.5 nm) [001] system with $\eta=0.14$. (c) Weighted spectrum on the surface of 9.9-nm-thick YIG with the outmost atomic layer removed. (d) The same as (a) with $\eta=1$.}
\label{fig4}%
\end{figure}
In Fig.{\ref{fig4}}(a), we project $|\mathcal{T}_{\alpha,{\bf k}}^{i'}|^{2}$ and $|\mathcal{T}_{\beta,{\bf k}}^{j'}|^{2}$ at GdIG surface to the hybrid spectrum in Fig.{\ref{fig2}}(a) with the ratio between the interfacial couplings $\eta=0.14$~\cite{geprags:nncomm2016}. Since the amplitudes of $\alpha_1$-type, $\alpha_2$-type and $\beta_2$-type modes at GdIG surface are sizable while those of $\beta_1$-type modes are rather small according to Fig.{\ref{fig2}}(b) and (c), the projections of $\alpha_1$-type, $\alpha_2$-type and $\beta_2$-type modes in Fig.{\ref{fig4}}(a) are much more visable than those of $\beta_1$-type modes. For similar reason, as shown in Fig.{\ref{fig4}}(b), the projections of $\beta_1$-type, $\beta_2$-type and $\alpha_2$-type magnons are relatively large at the YIG surface. Considering the negligible magnon occupation of $\alpha_2$-type magnons at low temperature, the SSCs in the YIG-GdIG-NM trilayer are mainly determined by $\alpha_1$-type and $\beta_2$-type modes while those in the GdIG-YIG-NM system are determined by $\beta_1$-type and $\beta_2$-type modes. Notice that the uniform mode in Fig.{\ref{fig4}}(b) does not contribute to the magnons on the YIG end. This is because we use an antiferromagnetic terminal plane in our calculation. In realistic situation, imperfect interface, different crystal orientations or the different coupling strengths, $\tilde{\mathcal{J}}_a$ and $\tilde{\mathcal{J}}_d$, will change the contribution of uniform mode and cause the measurable spin pumping signals~\cite{wangnioprl:2014}. To check the effect of imperfect interface on the uniform mode, we inspect a [001] orientated 8-unit-cell-thick (9.9 nm) YIG layer structure and find a ferromagnetic atomic plane under the outmost antiferromagnetic plane. We thus remove the topmost antiferromagnetic layer and see the acoustic mode has nonzero contribution as seen in Fig.{\ref{fig4}}(c). As $\eta$ is a free but crucial parameter, we increase $\eta$ to 1 in Fig.{\ref{fig4}}(d) and find the nearly dispersionless $\alpha_1$-type modes are enhanced more greatly than the others. This is because these low-frequency $\alpha_1$-type modes are dominated by the precession of Gd sublattice.

One of the most intriguing phenomena of LSSE in GdIG is the two sign-changing points (SCP) found in GdIG-NM bilayer system~\cite{geprags:nncomm2016}, where the higher and lower ones were attributed to the magnetic compensation and the competition between modes of opposite chiralites at the interface. As YIG-GdIG-NM trilayer owns the same interface as GdIG-NM bilayer, the lower SCP is also expected in YIG-GdIG-NM trilayer.

By solving the equation
\begin{align}\label{scp}
I_s={\rm \Delta T_{me}^{\beta}}(\gamma A_{\rm T}-B_{\rm T})=0,
\end{align}one can obtain the sign-changing temperature, which depends on two elements, i.e., the parameter $\gamma=(\Delta{\rm T_{mp}^{\alpha}}+{\rm \Delta T_{pp}^{\rm FN}})/({\rm \Delta T_{mp}^{\beta}+\Delta{\rm T_{pp}^{\rm FN}}})$ and the SSC. In general, both $\gamma$ and SSC could be function of YIG thickness: $\gamma$ is approximately 1 when YIG is very thin (just like the case in the GdIG-NM bilayer) and converge to a certain value when YIG is thick enough; SSC relies on the increase of YIG thickness due to the increase of subbands. Therefore, we study the relation between SCP and the YIG thickness with these two factors. From the discussion above, we see while the SSCs of hybrid structures with different YIG thicknesses can be calculated from the properties of magnons, but the value of $\gamma$ remains unclear. Here, we use a hypothetic function to describe $\gamma$ varying with the YIG thickness. Considering the smaller magnitude of $\Delta{\rm T_{mp}^{\alpha}}$ compared to $\Delta{\rm T_{mp}^{\beta}}$ according to the discussion at the beginning of this section, we assume
\begin{figure} [htpb]
\includegraphics [width=8.6cm]{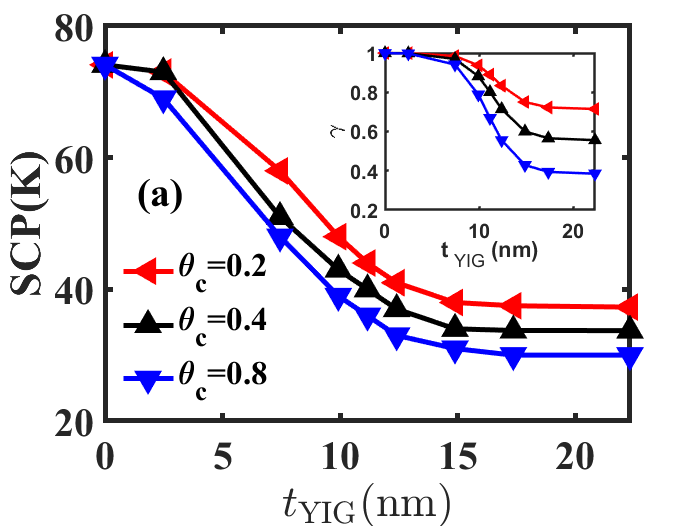}
\includegraphics [width=8.6cm]{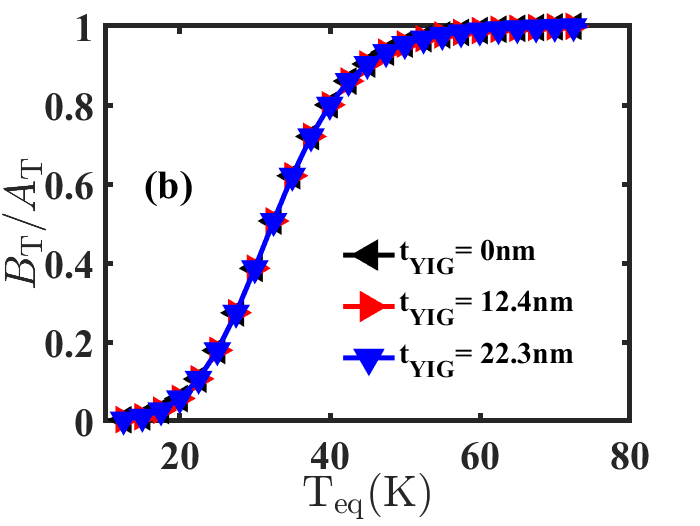}
\caption{(a) SCPs and $\gamma$ (inset) as function of YIG thickness in LSSE in YIG-GdIG-NM structure with different values of $\theta_c$. (b) The relation between the ratio of the two SSCs and ambient temperature for $t_{\rm YIG}=$ 0, 12.4 and 22.3 nm.}
\label{fig5}%
\end{figure}
\begin{align}
\Delta{\rm T_{mp}^{\alpha}}&=0.
\end{align}On the other hand, $\Delta{\rm T_{mp}^{\beta}}$ approximately equals to $\Delta{\rm T_{mp}^{\alpha}}$ if the YIG layer is very thin and significantly deviates from $\Delta{\rm T_{mp}^{\alpha}}$ when YIG becomes thick. We therefore use an asymptotic expression
\begin{align}\label{dtb}
\Delta{\rm T_{mp}^{\beta}} &= {\theta_c} \Delta{\rm T_{pp}^{FN}}\{\tanh[(t_{\rm YIG}-d_h)/\lambda_T]+ 1\},
\end{align}where $2\theta_c\Delta{\rm T_{pp}^{FN}}$ and $\lambda_T$ are the converged value and characteristic length of ${\rm \Delta T_{\rm mp}^{\beta}}$. $d_h$ is the thickness where $\Delta{\rm T_{mp}^{\beta}}$ reaches a half of converged value. When $d_h$ and $\lambda_T$ are set to be 10 and 2.5 unit cells respectively, the function in Eq.({\ref{dtb}}) at $t_{\rm YIG}=$ 0 nm and 22 nm gives $\Delta{\rm T_{mp}^{\beta}}=$ 0 and $2\theta_c \Delta{\rm T_{pp}^{FN}}$, respectively. Following this estimation, $\gamma$ can be expressed as

\begin{align}\label{SC}
 \gamma =\frac{1}{\theta_c\{\tanh[(t_{\rm YIG}-d_h)/\lambda_T]+ 1\}+1}.
\end{align}

For the value of $\theta_c$, we refer to the case in the YIG due to the lack of parameters in the GdIG.  Ref.\cite{schreier:prb2013} showed that the ratio between the magnon-phonon temperature difference and ${\rm \Delta T_{pp}^{FN}}$ is approximately 1 or 0.3 when the heat transfer between magnons in YIG and electrons in Pt is taken into account or not. We thus estimate $\theta_c$ = 0.2, 0.4, 0.8. Fig.{\ref{fig5}}(a) shows the SCPs as function of YIG thickness, where we find SCPs shift to lower temperature with the increase of YIG thickness by tens of Kelvins. To explain this feature, we refer to Eq.(\ref{scp}), which shows that at SCP, the ratio between the SSCs of $\beta$ and $\alpha$ modes equals to $\gamma$. The computational results for these ratios as function of temperature with different YIG thicknesses are shown in Fig.{\ref{fig5}}(b), revealing their negligible dependency on YIG thickness. Therefore, such large variation of SCPs are mainly caused by the change of $\gamma$. Note that the SCP for a given YIG thickness and $\theta_c$ is read from Fig.{\ref{fig5}}(b) by setting the ratio as the corresponding value of $\gamma$ in the inset of Fig.{\ref{fig5}}(a). According to a research in the heterostructure consisting of a ten-nm-thick garnet film and a normal metal layer\cite{leeanisotropictermprl:2020}, the interface might introduce an additional anisotropy due to the lattice mismatch and Rashba effect. We estimate such an anisotropic field could cause a correction to the frequency by only a few GHz, which is too small to affect our main results, dominated by the thermal magnons in THz range.

\section{Conclusion and Discussion}\label{sec5}
In summary, we study the properties of hybrid magnon modes in YIG-GdIG hybrid bilayer structure, which is naturally formed when YIG is grown on the substrate GGG. We find that the localized and extended features of different hybrid modes result in the distinct accumulations of magnons with opposite polarizations at surfaces. As magnons transfer spin angular momentum to electrons in an adjacent normal metal on GdIG side by magnon-electron scattering and thus cause nonzero spin current in the normal metal, we calculate this spin current in the longitudinal spin Seebeck configuration and recover a sign change in spin Seebeck signal, previously discovered in GdIG. More interestingly, we find the sign-changing temperature can vary by tens of Kelvins with the increase of YIG thickness.

\section{Acknowledgments}
This work was supported by the National Natural Science Foundation of China (Grant No.11974047) and Fundamental Research Funds for the Central Universities (Grant No. 2018EYT02). K. X. thanks the National Natural Science Foundation of China (Grants No. 61774017, No. 11734004) and NSAF (Grant No. U1930402).

\begin{appendix}
\section{Matrix elements in the bosonic BdG Hamiltonian}\label{parameter}
The matrix elements in Eq.(\ref{bosonic}) are defined as
  \begin{align}
  A_{ij}({\bf k})&=(2\sum_{\left|\boldsymbol{r}_{im}\right|=r_{aa}}J_{im}^{aa}S_{a,m}-2
  \sum_{\left|\boldsymbol{r}_{im}\right|=r_{ad}}J_{im}^{ad}S_{d,m}\notag\\
  &+2\sum_{\left|\boldsymbol{r}_{im}
  \right|=r_{ac}}J_{im}^{ac}S_{c,m})\delta_{ij}\notag\\
  &-2\sum_{\left|\boldsymbol{r}_{{ij}}\right|=r_{aa}}J_{ij}^{aa}\sqrt{S_{a,i}S_{a,j}}e^{i\boldsymbol{k}\cdot {\bf r}_{ij}},\notag \notag \\
  C_{ij}({\bf k})&=(2\sum_{\left|\boldsymbol{r}_{im}\right|=r_{cc}}J_{im}^{cc}S_{c,m}-\sum_{
  \left|\boldsymbol{r}_{im}\right|=r_{cd}}J_{im}^{cd}S_{d,m}\notag \\
  &+2\sum_{\left|\boldsymbol{r}_{im}\right|
  =r_{ac}}J_{im}^{ac}S_{a,m})\delta_{ij}\notag\\
  &-2\sum_{\left|\boldsymbol{r}_{{ij}}\right|=r_{cc}}J_{ij}^{cc}\sqrt{S_{c,i}S_{c,j}}e^{i\boldsymbol{k}\cdot {\bf r}_{ij}},\notag \\
  D_{ij}({\bf k})&=(2\sum_{\left|\boldsymbol{r}_{im}\right|=r_{dd}}J_{im}^{dd}S_{d,m}-2\sum_{
  \left|\boldsymbol{r}_{im}\right|=r_{ad}}J_{im}^{ad}S_{a,m}\notag \\
  &-2\sum_{\left|\boldsymbol{r}_{im}\right|=r_{cd}}
  J_{im}^{cd}S_{c,m})\delta_{ij}\notag\\
  &-2\sum_{\left|\boldsymbol{r}_{{ij}}\right|=r_{dd}}J_{ij}^{dd}\sqrt{S_{d,i}S_{d,j}}e^{i\boldsymbol{k}\cdot {\bf r}_{ij}},\notag \\
  B^{ss'}_{ij}({\bf k})&=-2\ensuremath{\sum_{\left|\boldsymbol{r}_{ij}
  \right|=r_{ss'}}J_{ij}^{ss'}\sqrt{S_{s,i}S_{s',j}}}e^{i\boldsymbol{k}\cdot\boldsymbol{r}_{ij}}.
  \end{align}
\section{Derivation of SSCs, $A_{\rm T}$ and $B_{\rm T}$}\label{appx}
 In this appendix, we derive the spin currents generated in the LSSE from the interfacial exchange Hamiltonian in Eq.(\ref{inth}). Assuming that the momentum conservation might be broken by roughness at the interface, we replace ${\bf q-k}$ by an independent vector ${\bf q'}$ in Eq.(\ref{inth}). Then we apply Fermi-Golden rule to calculate transition rates~\cite{shen:prb2016,miller:pr1960}
\begin{small}
\begin{align}\label{rate3}
\Gamma_{\uparrow\downarrow}&=\frac{2\pi}{\hbar}\mathcal{\tilde{J}}_{a}^{2}\sum_{{\bf q,k,q'}}[\sum_{i'}n_{i'}^{\text{{\bf k}}}\tau^{{\bf q}'{\bf q}}_{\uparrow\downarrow}|\mathcal{T}_{\alpha,{\bf k}}^{i'}|^{2}\delta(E_{\downarrow}^{{\bf q}}
-E_{\uparrow}^{{\bf q'}}-\hbar\omega_{i'}^{{\bf k}})\notag\\
&+\sum_{j'}(n_{j'}^{\text{-{\bf k}}}+1)\tau^{{\bf q}'{\bf q}}_{\uparrow\downarrow}|\mathcal{T}_{\beta,{\bf k}}^{j'}|^{2}\delta(E_{\downarrow}^{{\bf q}}-E_{\uparrow}^{{\bf q'}}+\hbar\omega_{j'}^{-{\bf k}})], \\
\Gamma_{\downarrow\uparrow}&=\frac{2\pi}{\hbar}\mathcal{\tilde{J}}_{a}^{2}\sum_{{\bf q,k,q'}}[\sum_{j'}n_{j'}^{\text{-{\bf k}}}\tau^{{\bf q}{\bf q}'}_{\downarrow\uparrow}|\mathcal{T}_{\beta,{\bf k}}^{j',\dagger}|^{2}\delta(E_{\uparrow}^{{\bf q'}}-E_{\downarrow}^{{\bf q}}-\hbar\omega_{j'}^{-{\bf k}})\notag\\
&+\sum_{i'}(n_{i'}^{\text{{\bf k}}}+1)\tau^{{\bf q}{\bf q}'}_{\downarrow\uparrow}|\mathcal{T}_{\alpha,{\bf k}}^{i',\dagger}|^{2}\delta(E_{\uparrow}^{{\bf q'}}-E_{\downarrow}^{{\bf q}}+\hbar\omega_{i'}^{{\bf k}})],
\end{align}
\end{small}where $\tau^{{\bf q}'{\bf q}}_{\uparrow\downarrow}=f_{\uparrow}^{{\bf q'}}(1-f_{\downarrow}^{{\bf q}})$ and $\tau^{{\bf q}{\bf q}'}_{\downarrow\uparrow}=f_{\downarrow}^{{\bf q}}(1-f_{\uparrow}^{{\bf q'}})$. $n_{j'}^{-{\bf k}}$ $(n_{i'}^{{\bf k}})$ is magnon distribution function for wave vector ${-\bf k}$ $({\bf k})$ and branch index $j' (i')$. $f_{\uparrow}^{\bf q}$ $(f_{\downarrow}^{{\bf q}'})$ is the distribution function of spin-up (spin-down) electrons of wave vector ${\bf q}$ $({\bf q}')$. Spin current is defined as the difference of these two processes
\begin{align}
I_{s}=\hbar(\Gamma_{\uparrow\downarrow}-\Gamma_{\downarrow\uparrow}).
\end{align}
Then we have the expression of spin current
\begin{small}
\begin{align}\label{spincurrent1}
I_s&=2\pi\mathcal{\tilde{J}}_{a}^{2}\sum_{{\bf k,q,q'}}\Big[\sum_{i'}n_{i'}^{\text{{\bf k}}}|\mathcal{T}_{\alpha,{\bf k}}^{i'}|^{2}\delta(E_{\downarrow}^{{\bf q}}-E_{\uparrow}^{{\bf q'}}-\hbar\omega_{i'}^{{\bf k}})\Delta f \notag\\
&+\sum_{j'}n_{j'}^{-{\bf k}}|\mathcal{T}_{\beta,{\bf k}}^{j'}|^{2}\delta(E_{\uparrow}^{{\bf q'}}\!-E_{\downarrow}^{{\bf q}}\!-\hbar\omega_{j'}^{-{\bf k}})\Delta f\Big]-X({\rm T_{e}}),
\end{align}
\end{small}where $\Delta f=f_{\uparrow}^{{\bf q'}}-f_{\downarrow}^{{\bf q}}$ and $X({\rm T_{e}})$ is the backflow from NM to magnetic insulator.

As the energy shifts of electrons are small compared to fermi energy, one has $f_{\uparrow}^{\bf q'}-f_{\downarrow}^{\bf q}\approx \partial_{E}f|_{E=E_{\bf q}}(E_{\uparrow}^{\bf q'}-E_{\downarrow}^{\bf q})$. At low temperature, $f(E)\approx \Theta(E_f-E)$, where $\Theta(E)$ is the Heaviside step function. Therefore $\partial_E f|_{E=E_{\bf q}}=-\delta(E_{\bf q}-E_f)$. When the transverse area is large enough so as to make wave vector quasi-continuous, the summation symbols of ${\bf q, q'}$ in Eq.(\ref{spincurrent1}) can be transformed to integral as $\sum_{{\bf q(q')}}A(E_{\bf q(q')})=\int\rho(E)A(E)dE$, where $\rho(E)$ is the density of states at energy E. The expression for spin current is therefore simplified into
\begin{small}
\begin{align}
I_{s}&=2\pi\hbar\rho(E_{f})\mathcal{\tilde{J}}_{a}^{2}\sum_{{\bf k}}[\sum_{i'}\omega_{i'}^{{\bf k}}n_{i'}^{\text{{\bf k}}}|\mathcal{T}_{\alpha,{\bf k}}^{i'}|^{2}\rho(E_{f}-\hbar\omega_{i'}^{{\bf k}})\notag \\
&-\sum_{j'}\omega_{j'}^{-{\bf k}}n_{j'}^{-{\bf k}}|\mathcal{T}_{\beta,{\bf k}}^{j',\dagger}|^{2}\rho(E_{f}+\hbar\omega_{j'}^{-{\bf k}})]-X({\rm T_{e}}).
\end{align}
\end{small}
Then the zero-order expression for spin current is
\begin{align}\label{spincurrent2}
I_{s}&\approx\mathcal{D}\hbar\sum_{{\bf k}}[\sum_{i'}\omega_{i'}^{{\bf k}}n_{i'}^{\text{{\bf k}}}|\mathcal{T}_{\alpha,{\bf k}}^{i'}|^{2}-\sum_{j'}\omega_{j'}^{-{\bf k}}n_{j'}^{-{\bf k}}|\mathcal{T}_{\beta,{\bf k}}^{j',\dagger}|^{2}]\notag\\
&-X({\rm T_{e}}),
\end{align}
where $\mathcal{D}=2\pi\rho^2(E_{f})\mathcal{\tilde{J}}_{a}^{2}$. When the system is in thermal equilibrium, no spin current is injected, which leads to
\begin{align}\label{xe}
X({\rm T_{eq}})&=\mathcal{D}\hbar\sum_{{\bf k}}[\sum_{i'}\omega_{i'}^{{\bf k}}n_{i'}^{\text{{\bf k}}}({\rm T_{eq}})|\mathcal{T}_{\alpha,{\bf k}}^{i'}|^{2}\notag \\
&-\sum_{j'}\omega_{j'}^{-{\bf k}}n_{j'}^{-{\bf k}}(\rm T_{eq})|\mathcal{T}_{\beta,{\bf k}}^{j',\dagger}|^{2}],
\end{align}where ${\rm T_{eq}}$ is the thermal equilibrium temperature (which should also be the ambient temperature). In near equilibrium, the temperature of electrons, ${\rm T_e}$, approximately equals to ${\rm T_{eq}}$. In this condition, we can substitute Eq.(\ref{xe}) into Eq.(\ref{spincurrent2}) and obtain the expressions for SSCs
\begin{align}\label{ssc}
A_{\rm T}&=\mathcal{D}\hbar\sum_{{\bf k}}\sum_{i'}(\partial_{{T}}n|_{T={\rm T_{eq}}})|\mathcal{T}_{\alpha,{\bf k}}^{i'}|^{2}\omega_{i'}^{{\bf k}},  \notag \\
B_{\rm T}&=\mathcal{D}\hbar\sum_{{\bf k}}\sum_{j'}(\partial_{{T}}n|_{T={\rm T_{eq}}})|\mathcal{T}_{\beta,{\bf k}}^{j',\dagger}|^{2}\omega_{j'}^{{\bf -k}}.
\end{align}
\end{appendix}
\end{document}